\documentclass[aoas,MSNbibl,nameyear,dvips]{arximspdf}
\usepackage{graphicx}

\doi{10.1214/11-AOAS463}
\volume{5}
\issue{3}
\pubyear{2011}
\firstpage{1978}
\lastpage{2002}

\makeatletter
\def\thick{\bolds}
\def\bsuffix #1{#1}

  \let\sv@tabnotetext\tabnotetext
  \let\sv@tabnotemark@fmt\tabnotemark@fmt
   \long\def\legend#1{{\let\tabnote@indent\leavevmode\sv@tabnotetext[]{}{#1}}}

\def\bptnote#1{}
\makeatother

\begin{document}
\begin{frontmatter}

\title{Incorporating biological information into linear models: A Bayesian approach
    to the selection of pathways and genes}
\runtitle{A Bayesian model for pathway and gene selection}

\begin{aug}
\author[A]{\fnms{Francesco C.} \snm{Stingo}\ead[label=e1]{fcs1@rice.edu}},
\author[B]{\fnms{Yian A.} \snm{Chen}\ead[label=e2]{Ann.Chen@moffitt.org}},
\author[C]{\fnms{Mahlet G.} \snm{Tadesse}\ead[label=e3]{mgt26@georgetown.edu}}\\
\and
\author[A]{\fnms{Marina} \snm{Vannucci}\corref{}\thanksref{m5}\ead[label=e4]{marina@rice.edu}}
\runauthor{Stingo, Chen, Tadesse and Vannucci}
\affiliation{Rice University, Moffitt Cancer Center,
Georgetown University and~Rice~University}
\address[A]{F. C. Stingo\\
M. Vannucci\\
Department of Statistics\\
Rice University\\
Houston, Texas
77005\\
USA\\
\printead{e1}\\
\textsc{E-mail:}\ \printead*{e4}} 
\address[B]{Y. A. Chen\\
Moffitt Cancer Center\\Tampa, Florida 33612\\ USA\\
\printead{e2}}
\address[C]{M. G. Tadesse\\
Department of Mathematics and Statistics\\ Georgetown University\\
Washington, DC 20057\\ USA\\
\printead{e3}}
\end{aug}
\thankstext{m5}{Supported in part by NIH Grant
R01-HG0033190-05 and NSF Grant DMS-10-07871.}

\received{\smonth{10} \syear{2009}}
\revised{\smonth{2} \syear{2011}}

%
\begin{abstract}
The vast amount of biological knowledge accumulated over the years has
allowed researchers to identify various biochemical
interactions and define different families of pathways. There is an
increased interest in identifying pathways and pathway
elements involved in particular biological processes. Drug discovery
efforts, for example, are focused on identifying
biomarkers as well as pathways related to a disease. We propose a
Bayesian model that addresses this question by
incorporating information on pathways and gene networks in the analysis
of DNA microarray data. Such information is used to
define pathway summaries, specify prior distributions, and structure
the MCMC moves to fit the model. We illustrate the
method with an application to gene expression data with censored
survival outcomes. In addition to identifying markers that
would have been missed otherwise and improving prediction accuracy, the
integration of existing biological knowledge into
the analysis provides a better understanding of underlying molecular processes.
\end{abstract}

%
\begin{keyword}
\kwd{Bayesian variable selection}
\kwd{gene expression}
\kwd{Markov chain Monte Carlo}
\kwd{Markov random field prior}
\kwd{pathway selection}.
\end{keyword}

\end{frontmatter}
\setcounter{footnote}{1}

\section{Introduction}\label{sec1}

DNA microarrays have been used successfully to identify gene expression
signatures characteristic of disease subtypes
[\citet{golub1999}] or distinct outcomes to therapy [\citet
{shipp2002}]. Many statistical methods have
been developed to select genes for disease diagnosis, prognosis and
therapeutic targets. However, gene selection alone may not be
sufficient. In cancer pharmacogenomics, for instance, cancer drugs are
increasingly designed to target specific pathways to account for the
complexity of the oncogenic process and the complex
relationships between genes [\citet{downward2006}]. Metabolic
pathways, for example, are defined as a series of
chemical reactions in a living cell that can be activated or inhibited
at multiple points. If a gene at the
top of a signaling cascade is selected as a target, it is not
guaranteed that the reaction will be successfully inactivated,
because multiple genes downstream can still be activated or inhibited.
Signals are generally relayed via multiple signaling
routes or networks. Even if a branch of the pathway is completely
blocked by inhibition or activation of multiple genes, the
signal may still be relayed through an alternative branch or even
through a different pathway
[\citet{bild2006}]. \citet{downward2006} pointed out that targeting a
single pathway or a few
signaling pathways might not be sufficient. Thus, the focus is
increasingly on identifying both relevant genes and
pathways. Genes and/or gene products generally interact with one
another and they often function together concertedly. Here we propose a
Bayesian model that addresses this question by incorporating
information of pathway memberships and
gene networks in the analysis of DNA microarray data. Such information
is used to define pathway summaries, specify prior
distributions, and structure the MCMC moves.

Several public and commercial databases have been developed to
structure and store the vast amount of biological knowledge
accumulated over the years into functionally or biochemically related
groups. These databases focus on describing signaling,
metabolic or regulatory pathways. Some examples include Gene Ontology
(GO) [\citet{go2000}], Kyoto Encyclopedia of
Genes and Genomes (KEGG) [\citet{kegg2000}], MetaCyc [\citet
{krieger2004}], PathDB
, Reactome
KnowledgeBase [\citet{reactome2005}], Invitrogen iPath
(\href{http://www.invitrogen.com}{www.invitrogen.com}) and Cell Signaling Technology (CST) Pathway
(\href{http://www.cellsignal.com}{www.cellsignal.com}). The need to integrate gene expression data with
the biological knowledge accumulated in these
databases is well recognized. Several software packages that query
pathway information and overlay DNA microarray data on
pathways have been developed. \citet{nakao1999} implemented a
visualization tool that color-codes KEGG pathway diagrams to
reflect changes in their gene expression levels. GenMAPP [\citet
{dahlquist2002}] is another graphical tool that allows
visualization of microarray data in the context of biological pathways
or any other functional grouping of genes.
\citet{doniger2003} use GenMAPP to view genes involved in specific GO
terms. Another widely used method that relates pathways to a set of
differentially expressed genes is the gene set enrichment analysis (GSEA)
[\citet{Subramanian2005}]. Given a list of genes, GSEA computes an
enrichment score to reflect the degree to which a predefined pathway is
over-represented at the top or bottom of the ranked list. These
procedures are useful starting points to observe gene expression
changes for known biological processes.

Recent studies have gone a step further and focused on incorporating
pathway information or gene--gene network
information into the analysis of gene expression data. For example,
\citet{park2007} have attempted to incorporate GO
annotation to predict survival time, first grouping genes based on
their GO membership, calculating the first principal
component to form a super-gene within each cluster and then applying a
Cox model with $L_1$ penalty to identify super-genes,
that is, GO terms related to the outcome. \citet{wei2007} have
considered a small set of 33 preselected signaling pathways
and used the implied relationships among genes to infer differentially
expressed genes, and \citet{wei2008} have extended
this work by including a temporal dimension. \citet{li2008} and
\citet{pan2009} have proposed different procedures
that use the gene--gene network to build penalties in a regression model
for gene selection. Bayesian approaches
have also been developed. \citet{zhang2010} have incorporated the
dependence structure of transcription factors in a
regression model with gene expression outcomes. There, a network is
defined based on the Hamming distance between
candidate motifs and used to specify a Markov random field prior for
the motif selection indicator. \citet{donatello2008}
have proposed a model for the identification of differentially
expressed genes that takes into account the dependence
structure among genes from available pathways while allowing for
correction in the gene network topology.
\citet{stingo2011} use a Markov random field prior that captures the
gene--gene interaction network in a discriminant
analysis setting.

These methods use the gene-pathway relationships or gene network
information to identify either the important pathways or
the genes. Our goal is to develop a more comprehensive method that
selects both pathways and genes using a model that
incorporates pathway-gene relationships and gene dependence structures.
In order to identify relevant genes and pathways,
latent binary vectors are introduced and updated using a two-stage
Metropolis--Hastings sampling scheme. The gene networks
are used to define a Markov random field prior on the gene selection
indicators and to structure the Markov chain Monte
Carlo (MCMC) moves. In addition, the pathway information is used to
derive pathway expression measures that summarize the
group behavior of genes within pathways. In this paper we make use of
the first
latent components obtained by applying partial least squares (PLS)
regressions on the selected genes from each pathway. PLS is an
efficient statistical regression technique that was initially proposed
in the chemometrics literature [\citet{wold}] and more recently used
for the analysis
of genomic and proteomic data; see \citet{strimmer}. We
apply the model to simulated and real data using the pathway structure
from the KEGG database.

Our simulation studies show that the MRF prior leads to a better
separation between relevant and nonrelevant pathways,
and to less false positives in a model with fairly small regression
coefficients.
Other authors have reported similar results. \citet{zhang2010}, in
particular, comment on the effect of
the MRF prior on the selection power in their linear regression
setting. They also notice that adding the MRF prior implies a
relatively small increase in computational cost. \citeauthor{wei2007}
(\citeyear{wei2007,wei2008}) report that their method is quite effective in
identifying genes and modified subnetworks and that it has higher
sensitivity than commonly used procedures that do not use the
pathway structure, with similar and, in some cases, lower false
discovery rates. Furthermore, in our model formulation we use the
network information not only for prior specification but also to
structure the MCMC moves. This is helpful for
arriving at promising models faster by proposing relevant
configurations. In real data applications the
integration of pathway information may allow the identification of
relevant predictors that could be missed
otherwise, aiding the interpretation of the results, in particular, for
the selected genes that are connected in the MRF, and also improving
the prediction accuracy of selected models.

The paper is organized as follows. Section~\ref{sec2} contains the model
formulation and prior specification. Section~\ref{sec3}
describes the MCMC procedure and strategies for posterior inference. In
Section~\ref{sec4} performances are evaluated on simulated data and an
application of the method to gene expression data with survival outcomes is presented.
Section~\ref{sec5} concludes the paper with a~brief discussion.

\section{Model specification}\label{sec:model}\label{sec2}
\begin{figure}

\includegraphics{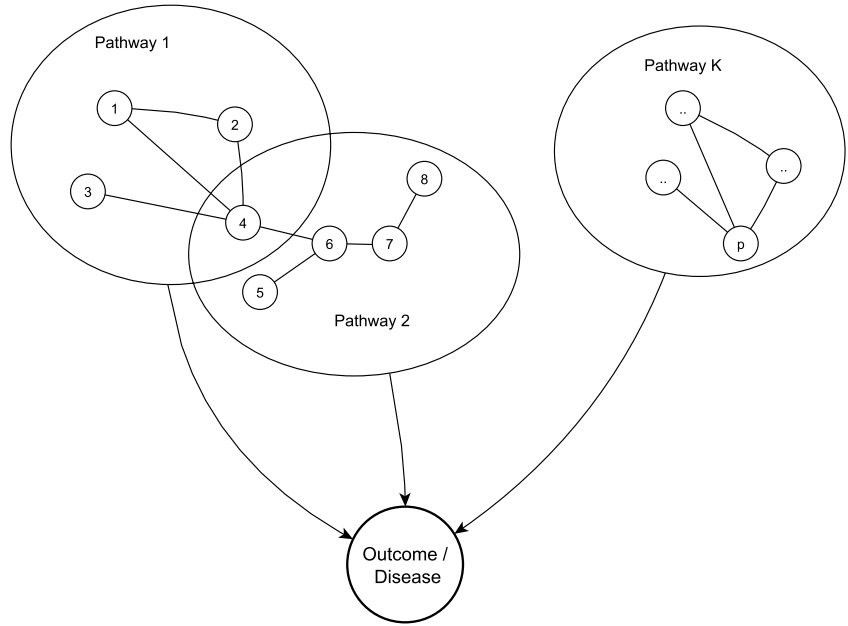}

\caption{Schematic representation of our proposed approach.
Information on known pathways and gene--gene networks is obtained from
available databases. PLS components restricted to known pathways serve
as possible regressors to predict a disease outcome, according to model
(\protect\ref{eq:model2}). The goal of the inference is to identify the
pathways to be included in the model and the genes to be included
within those pathways.}
\label{figure1}
\end{figure}

We describe how we incorporate pathway and gene network information
into a Bayesian modeling framework for
gene and pathway selection. Figure \ref{figure1} represents a
schematic representation of our approach and model.

\subsection{Regression on latent measures of pathway activity}\label{sec:reg}\label{sec2.1}
Our goal is to build a model for identifying pathways related to a
particular phenotype while simultaneously locating genes
from these selected pathways that are involved in the biological
process of interest. The data we have available for
analysis consist of the following:
\begin{longlist}[(4)]
\item[(1)]$Y$, an $n \times1$ vector of outcomes.
\item[(2)]$\mathbf{X}$, an $n \times p$ matrix of gene expression levels.
Without loss of generality, $\mathbf{X}$~is centered so that its
columns sum to 0.
\item[(3)]$\mathbf{S}$, a $K \times p$ matrix indicating membership of genes
in pathways, with elements $s_{kj}=1$ if gene $j$ belongs
to pathway $k$, and $s_{kj}=0$ otherwise.
\item[(4)]$\mathbf{R}$, a $p \times p$ matrix describing relationships between
genes, with $r_{ij}=1$ if genes $i$ and $j$
have a direct link in the gene network, and $r_{ij}=0$ otherwise.
\end{longlist}
The matrices $\mathbf{S}$ and $\mathbf{R}$ are constructed using information
retrieved from pathway databases; see the application in Section
\ref{sec:app} for details.

Since the goal of the analysis is to study the association between the
response variable and the pathways, we need to derive
a score as a measure of ``pathway expression'' that summarizes the
group behavior of included genes within pathways. We
do this by using the latent components from a PLS regression of $Y$ on
selected subsets of genes from each pathway. A
number of recent studies have, in fact, applied dimension reduction
techniques to capture the group behavior of multiple
genes. \citet{pittman2004}, for instance, first apply $k$-means
clustering to identify subsets of potentially related
genes, then use as regressors the first principal components obtained
from applying principal component analysis (PCA) to
each cluster. \citet{bair2006} start by removing genes that have low
univariate correlation with the outcome variable, then
apply PCA on the remaining genes to form clusters or conceptual
pathways, which are used as regressors. In our method,
instead of attempting to infer conceptual pathways, we use the existing
pathway information. We compute a pathway activity
measure by applying PLS regression of $Y$ on a subset of selected genes
from the pathway. PLS has the advantage of taking
into account the covariance between regressors and the response
variable~$Y$, whereas PCA focuses solely on the variability
in the covariate data. The selection of a subset of gene expressions to
form the PLS components is similar in spirit to the
sparse PCA method proposed by \citet{zou}, which selects variables to
form the principal components.

To identify both relevant groups and important genes, we introduce two
binary vector indicators, a $K \times1$
vector $\thick\theta$ for the inclusion of the groups and a $p \times
1$ vector $\thick\gamma$ for the inclusion of genes, that is,
$\gamma_j=1$ if gene $j$ is selected for at least one pathway score,
and $\gamma_j=0$ otherwise. Assuming that the response $Y$
is continuous, the linear regression model that relates $Y$ to the
selected pathways
and genes is
\begin{equation}\label{eq:model2}
Y = {\mathbf1} \alpha+ \sum_{k=1}^{K_\theta} T_{k (\gamma)}
\beta_{k (\gamma)} + \thick\varepsilon,   \qquad
\thick\varepsilon\sim\mathcal{N}({\mathbf0},\sigma^2 \mathbf{I}),
\end{equation}
where $K_\theta= \sum_{k=1}^K \theta_k$ is the number of selected
pathways and where $T_{k (\gamma)}$ corresponds to the
first latent PLS component generated based on the expression levels of
selected genes belonging to pathway $k$, that is,
using the~$X_j$'s corresponding to $s_{kj}=1$ and $\gamma_j=1$. To be
more precise, let pathway~$k$ contain $p_k =
\sum_{j=1}^p s_{kj}$ genes and let $p_{k \gamma} = \sum_{j=1}^p
s_{kj} \gamma_j$ denote the number of selected genes
(i.e., genes included in the model) that belong to pathway~$k$. Then
$T_{k(\gamma)}$ corresponds to the first latent PLS
component generated by applying PLS to the expression data of the $p_{k
\gamma}$ genes, denoted as $\mathbf{X}_{k(\gamma)}$,
\[
T_{k(\gamma)} = \mathbf{X}_{k(\gamma)} U_1,
\]
where $U_1$ is the $p_{k\gamma}\times1$ eigenvector corresponding to
the largest eigenvalue of $C_{xy}C_{xy}^T$, with
$C_{xy} = \operatorname{cov}(\mathbf{X}_{k(\gamma)},Y )$ [see, e.g., \citet
{lindgren1993}]. Thus, $T_k(\gamma)$ is an $n \times1$ vector and
$\beta_k(\gamma)$ is a
scalar. Model (\ref{eq:model2}) can therefore be seen as a PLS
regression model with PLS components restricted to available
pathways, and where the goal of the inference is to identify the
pathways to be included in the model, and the genes to be
included within those pathways.

\subsection{Models for categorical or censored outcomes} \label{sec:surv}\label{sec2.2}
In the construction above, we have assumed a continuous response.
However, our model formulation can easily be extended to
handle categorical or censored outcome variables.

When $Y$ is a categorical variable taking one of $G$ possible values,
$0, \ldots,\allowbreak G-1$, a probit model can be used, as done
by \citet{albert1993}, \citet{sha2004} and \citet{deukwoo2007}.
Briefly, each outcome $Y_i$ is associated with a vector
$(p_{i,0}, \ldots, p_{i,G-1})$, where $p_{ig} = P(Y_i=g)$ is the
probability that subject $i$ falls in the $g$th category.
The probabilities $p_{ig}$ can be related to the linear predictors
using a data augmentation approach. Let $\mathbf{Z}_i$ be
latent data corresponding to the unobserved propensities of subject $i$
to belong to one of the classes. When the observed
outcomes $Y_i$ correspond to nominal values, the relationship between
$Y_i$ and $\mathbf{Z}_i = (z_{i,1}, \ldots, z_{i,G-1})$
can be defined~as
\begin{equation}
\label{eq:probit}
Y_i =
\cases{\displaystyle  0 ,&\quad   if    $\displaystyle\max_{1 \le l \le G-1} \{
z_{i,l}\} \le
0$,\vspace*{2pt}\cr
\displaystyle g ,&\quad   if     $\displaystyle\max_{1 \le l \le G-1} \{z_{i,l}\} > 0$
 and    $\displaystyle z_{i,g} = \max_{1 \le l \le G-1} \{ z_{i,l}\}$.}
\end{equation}
A multivariate normal model can then be used to associate $\mathbf{Z}_i$
to the predictors
\begin{equation}
\label{eq:augment1}
\mathbf{Z}_i = {\mathbf1}\alpha+ \sum_{k=1}^{K_\theta} T_{i, k (\gamma)}
\thick\beta_{k(\gamma)} + \thick\varepsilon_i ,
 \qquad  \thick\varepsilon_i \sim\mathcal{N}(0,\thick\Sigma),
i=1, \ldots, n.
\end{equation}

If the observed outcomes $Y_i$ correspond, instead, to ordinal
categories, the latent variable $Z_i$ is defined such that
$Y_i = g$     if     $\delta_g < Z_i \le\delta_{g+1}$,
$g=0,\ldots, G-1$,
where the boundaries $\delta_g$ are unknown and $-\infty= \delta_0 <
\delta_1 < \cdots < \delta_{G-1} < \delta_G = \infty$.
The latent variable $Z_i$ is associated with the predictors through the
linear model
%
\begin{equation}\label{eq:augment2}
Z_i = \alpha+ \sum_{k=1}^{K_\theta} T_{i, k (\gamma)} \beta_{k
(\gamma)} + \varepsilon_i ,  \qquad
\varepsilon_i \sim\mathcal{N}(0,\sigma^2),   i=1, \ldots, n.
\end{equation}

For censored survival outcomes, an accelerated failure time (AFT) model
can be used [\citet{wei1992}; \citet{sha2006}]. In this case, the observed
data are $Y_i = \min(T_i, C_i)$ and $\delta_i = I\{Y_i \le C_i\}$,
where $T_i$
is the survival time for subject $i$, $C_i$ is the censoring time, and
$\delta_i$ is a censoring indicator. A data
augmentation approach can be used and latent variables~$Z_i$ can be
introduced such that
\begin{equation}\label{eq:survival}
\cases{\displaystyle
Z_i = \log(Y_i) ,&\quad  if    $\delta_i = 1$,\cr\displaystyle
Z_i > \log(Y_i) ,&\quad  if    $\delta_i = 0$.
}
\end{equation}
The AFT model can then be written in terms of the latent $Z_i$
similarly to (\ref{eq:augment2})
where the $\varepsilon_i$'s are independent and identically
distributed random variables that may take one of several
parametric forms. \citet{sha2006} consider cases where $\varepsilon
_i$ follows a normal or a $t$-distribution.

\subsection{Prior for regression parameters} \label{sec:priorbeta}
The regression coefficient $\beta_{k}$ in (\ref{eq:model2}) measures
the effect of the PLS latent component summarizing the
effect of pathway $k$ on the response variable. However, not all
pathways are related to the phenotype and the goal is to
identify the predictive ones. Bayesian methods that use mixture priors
for variable selection have been thoroughly
investigated in the literature, in particular, for linear models; see
\citet{george1997} for multiple
regression, \citet{brown1998} for extensions to multivariate responses
and \citet{sha2004}\vadjust{\eject} for probit models. A
comprehensive review on features of the selection priors and on
computational aspects of the method can be found in
\citet{chipman}. Similarly, we use the latent vector $\thick\theta$
to specify a scale mixture of a normal density
and a point mass at zero for the prior on each~$\beta_{k}$ in (\ref
{eq:model2}):\looseness=-1
\begin{equation} \label{eq:priorbeta}
\beta_k|\theta_k, \sigma^2 \sim\theta_k \cdot\mathcal{N}(\beta
_0,h \sigma^2) + (1-\theta_k) \cdot\delta_0(\beta_k),
 \qquad  k=1,\ldots, K,
\end{equation}\looseness=0
where $\delta_0(\beta_k)$ is a Dirac delta function. The
hyperparameter $h$ in (\ref{eq:priorbeta}) regulates, together with
the hyperparameters of $p(\thick\theta, \thick\gamma| \eta)$
defined in Section~\ref{sec2.4} below, the amount of shrinkage in the model. We
follow the
guidelines provided by \citet{sha2004} and specify $h$ in the range of
variability of the data so as to control the ratio of
prior to posterior precision.
For the intercept term, $\alpha$, and the variance, $\sigma^2$, we
take conjugate priors
$\alpha|\sigma^2 \sim \mathcal{N}(\alpha_0,h_0 \sigma^2)$ and
$\sigma^2 \sim  \operatorname{Inv\mbox{-}Gamma}(\nu_0/2,\nu_0
\sigma_0^2/2)$,
where $\alpha_0$, $\beta_0$, $h_0$, $h$, $\nu_0$ and $\sigma^2_0$
are to be elicited.

\subsection{Priors for pathway and gene selection indicators}\label{sec:priorSelection}\label{sec2.4}
In this section we define the prior distributions for the pathway
selection indicator, $\thick\theta$, and gene selection
indicator, $\thick\gamma$. These priors are first defined marginally
then jointly, taking into account some necessary
constraints.

Each element of the latent $K$-vector $\thick\theta$ is defined as
%
\begin{equation}
\theta_k =
\cases{\displaystyle
1 ,&\quad  if pathway $k$ is represented in the model,\cr\displaystyle
0 ,&\quad  otherwise}
\end{equation}
for $k=1,\ldots,K$. We assume independent Bernoulli priors for the
$\theta_k$'s,
%
\begin{equation}
p(\thick\theta|\varphi_k) = \prod_{k=1}^K \varphi_k^{\theta_k}
(1-\varphi_k)^{1-\theta_k},
\end{equation}
where $\varphi_k$ determines the proportion of pathways expected
 a priori  in the model. A mixture prior can be
further specified for $\varphi_k$ to achieve a better discrimination
in terms of posterior probabilities between significant
and nonsignificant pathways by inflating $p(\theta_k = 0)$ toward 1
for the nonrelevant pathways, as first suggested by
\citet{lucas2006},
\begin{equation}
p(\varphi_k) = \rho\delta_0(\varphi_k) + (1 - \rho)
\mathcal{B}(\varphi_k | a_0,b_0),
\end{equation}
where $\mathcal{B}(\varphi_k | a_0,b_0)$ is a Beta density function
with parameters $a_0$ and $b_0$. Since inference on
$\varphi_k$ is not of interest, it can be integrated out to simplify
the MCMC implementation. This leads to the following
marginal prior for~$\thick\theta$:
\begin{equation} \label{eq:priorphi}
p(\thick\theta) = \prod_k \biggl [ \rho\cdot(1-\theta_k) +
(1-\rho) \cdot\frac{B(a_0+\theta_k, b_0+1-\theta_k)}{B(a_0,b_0)}
 \biggr],
\end{equation}
where $B(\cdot,\cdot)$ is the Beta function. Prior (\ref
{eq:priorphi}) corresponds to a product of Bernoulli
distributions with parameter $\varphi_k^* =
\frac{a_0(1-\rho)}{a_0+b_0}$.\vadjust{\eject}

For the latent $p$-vector $\thick\gamma$ we specify a prior
distribution that is able to take into account not only the
pathway membership of each gene but also the biological relationships
between genes within and across pathways, which are
captured by the matrix $R$. Following \citet{zhang2010}, we model
these relations using a Markov random field (MRF), where
genes are represented by nodes and relations between genes by edges. A
MRF is a graphical model in which the distribution of
a set of random variables follow Markov properties that can be
described by an undirected graph. In particular, two unconnected
genes are considered conditionally independent given all other genes
[\citet{besag}]. Relations on
the MRF are represented by the following probabilities:
\begin{equation}
p(\gamma_j|\eta, \gamma_i, i \in N_j) = \frac{\exp(\gamma_j
F(\gamma_j))}{1+\exp(F(\gamma_j))}, \label{MRF}
\end{equation}
where $F(\gamma_j)= (\mu+\eta\sum_{i \in N_j}\gamma_i))$ and $N_j$
is the set of direct neighbors of gene $j$ in the MRF
using only pathways represented in the model,  that is, pathways
with $\theta_k=1$. The corresponding global distribution
on the MRF is given by
\begin{equation}
\label{PriorMRF}
p(\thick\gamma|{\thick\theta},\mu,\eta) \propto\exp(\mu\mathbf
1_p' \thick\gamma+ \eta\thick\gamma' \mathbf{R} \thick\gamma),
\end{equation}
with $\mathbf1_p$ the unit vector of dimension $p$ and $\mathbf{R}$ the
matrix introduced in Section \ref{sec:reg}. The parameter $\mu$
controls the
sparsity of the model, while $\eta$ regulates the smoothness of the
distribution of ${\thick\gamma}$ over the graph by controlling the
prior probability of selecting a gene based on how many of its
neighbors are selected. In particular, higher values of $\eta$
encourage the selection of genes with neighbors already selected into
the model. If a gene does not have any neighbor, then its prior
distribution reduces to an independent Bernoulli with
parameter $p = \exp(\mu) /[1+\exp(\mu)]$, which is a logistic
transformation of $\mu$.

Here, unlike \citet{zhang2010}, who fix both parameters of the MRF
prior, we
specify a hyperprior for $\eta$. We give positive probability to
values of~$\eta$ bigger than $0$,
which is biologically more intuitive than negative values of this
parameter (which would favor neighboring genes to have different
inclusion status). Such restriction on the domain of $\eta$ also
minimizes the ``phase transition'' problem that typically occurs with
MRF parameterizations of type (\ref{MRF}), where the
dimension of the selected model increases massively for small
increments of $\eta$.
When the phase transition occurs the number of selected genes increases
substantially. Here, after having detected the phase transition value
$\eta_{PT}$, by simulating from (\ref{PriorMRF}) over a grid of $\eta
$ values,
we specify a Beta distribution $\operatorname{Beta}(c_0,d_0)$ on $\eta/\eta_{PT}$.

Constraints need to be imposed to ensure both interpretability and
identifiability of the model. We essentially want to avoid the following:
\begin{longlist}[(3)]
\item[(1)] empty pathways,   that is, selecting a pathway but none of
its member genes;\vadjust{\eject}
\item[(2)] orphan genes, that is, selecting a gene but none of the
pathways that contain it;
\item[(3)] selection of identical subsets of genes by different pathways, a
situation that generates identical values
$T_{k(\gamma)}$ and $T_{k'(\gamma)}$ to be included in the model.
\end{longlist}
These constraints imply that some combinations of $\thick\theta$ and
$\thick\gamma$ values are not allowed. The joint prior probability
for $(\thick\theta, \thick\gamma)$ taking into account these
constraints is given by
\[ \label{priorTG}
p(\thick\theta, \thick\gamma| \eta) \propto
\cases{\displaystyle
\prod_{k=1}^K \varphi_k^{*\theta_k} (1-\varphi_k^*)^{1-\theta_k}
\exp(\mu\mathbf 1_p' \thick\gamma+ \eta\thick
\gamma' \mathbf{R} \thick\gamma) ,\cr\qquad  \mbox{for valid configurations,}\cr
0 , \quad  \mbox{for invalid configurations.}
}
\]

\section{Model fitting}\label{sec:mcmc}\label{sec3}

We now describe our MCMC procedure to fit the model and discuss
strategies for posterior inference with huge
posterior spaces, as in this model. In the Bayesian literature on
variable selection
for standard linear regression models stochastic search algorithms have
been designed to explore the posterior space, and
have been successfully employed in genomic applications with
prohibitive settings, handling models with thousands of genes. A key to
these applications is the assumption of sparsity of the model, that is,
the belief that the response is associated with a
small number of regressors. A stochastic search then allows one to
explore the posterior space in an effective way, quickly
finding the most probable configurations, that is, those corresponding
to coefficients with high marginal probabilities,
while spending less time in regions with low posterior probability.

We describe below the MCMC algorithm we have designed for our problem.
In particular, borrowing from the literature on
stochastic searches for variable selection, we work with a marginalized
model and design a~Metro\-polis--Hastings algorithm that updates the indicator
parameters for the inclusion of pathways and genes with a set of moves
that add and/or delete a single gene and a single
pathway. Also, we update the parameter $\eta$ of the MRF from its
posterior distribution by employing the general method
proposed by \citet{moller2006}. In \citet{supplmat} we discuss how our
Bayesian stochastic search variable selection kernel
generates an ergodic Markov chain over the restricted space. In
applications, we have found that a good way to asses if the
stochastic exploration can be considered satisfactory is to check the
concordance of the posterior probabilities obtained
from different chains started from different initial points.

\subsection{Marginal posterior probabilities}\label{sec3.1}
The model parameters consist of $(\alpha, \thick\beta,$ $\sigma^2,
{\thick\gamma}, \thick\theta, \eta)$. The MCMC procedure can
be made more efficient by integrating out some of the parameters.
Here,\vadjust{\eject}
we integrate out the regression parameters,
$\alpha$, $\thick\beta$ and $\sigma^2$. This leads to a
multivariate $t$-distribution
\begin{equation} \label{eq:fulcondy}
 \hspace*{24pt} f(Y|\mathbf  T, \thick\theta, \thick\gamma) \!\sim\!
\mathcal{T}_{\nu_0}\bigl(\alpha_0 \mathbf{1}_n \!+\! \mathbf  T_{(\theta,
\gamma)} \beta_0, \sigma_0^2 \bigl(\mathbf{I}_n \!+\! h_0
\mathbf{1}_n\mathbf{1}_n' \!+\!\mathbf  T_{(\theta, \gamma)} \thick\Sigma_0
\mathbf  T_{(\theta, \gamma)}'\bigr) \bigr),
\end{equation}
with $\nu_0$ degrees of freedom and $\mathbf{1}_n$ an $n$-vector of
ones, and where $\thick\Sigma_0= h \mathbf{I}_{K_{\theta}}$,
with $\mathbf{I}_n$ the $n \times n$ identity matrix, and ${\mathbf
T}_{(\theta, \gamma)}$ the $n \times K_{\theta}$ matrix derived
from the first PLS latent components for the selected pathways using
the selected genes. In the notation (\ref{eq:fulcondy}) the two
arguments of the $t$-distribution represent the mean and the scale
parameter of the distribution, respectively. The
posterior probability distribution of the pathway and gene selection
indicators is then given by
\begin{equation}\label{fullcond}
f(\thick\theta,\thick\gamma, \eta|\mathbf  T, Y) \propto f(Y|\mathbf
T,\thick\theta, \thick\gamma) \cdot p( \thick\theta, \thick
\gamma|\eta) \cdot p(\eta).
\end{equation}

\subsection{MCMC sampling}\label{sec3.2}
The MCMC steps consist of the following: (I) sampling pathway and gene
selection indicators from $p(\thick\theta,
\thick\gamma| \mbox{rest})$; (II) sampling the MRF parameter from
$p(\eta|\mbox{rest})$; (III) sampling additional
parameters introduced when fitting probit models for categorical
outcomes or AFT models for survival data.
\begin{enumerate}[(III)]
\item[(I)] The parameters $(\thick\theta, \thick\gamma)$ are
updated using a Metropolis--Hastings algorithm in a two-stage
sampling scheme. The pathway-gene relationships are used to structure
the moves and account for the constraints specified
in Section \ref{sec:priorSelection}. Details of the MCMC moves to
update $(\thick\theta, \thick\gamma)$ are given in \citet
{supplmat}. They consist of randomly choosing one of the following move types:

\begin{enumerate}[(3)]
\item[(1)]  change the inclusion status of gene and pathway by randomly
choosing between adding a pathway and a
gene or removing them both;
\item[(2)]  change the inclusion status of gene but not pathway by
randomly choosing between adding a gene or removing a gene;
\item[(3)]  change the inclusion status of pathway but not gene by
randomly choosing between adding a pathway or removing a
pathway.
\end{enumerate}
\item[(II)] At this step we want to draw the MRF parameter $\eta$
from the posterior density
$p(\eta|\thick\gamma) \propto p(\eta) p(\thick\gamma|\eta)$.
The prior distribution on $\thick\gamma$ is of the form
%
\begin{equation}
\label{moller1}
p(\thick\gamma|\eta) = q_{\eta} (\thick\gamma) / Z_{\eta}
\end{equation}
with unnormalized density $q_{\eta} (\thick\gamma)$ and a
normalizing constant $Z_{\eta}$ which is not available analytically.
When calculating the Metropolis--Hastings ratio to determine the
acceptance probability of a new value $\eta^{p}$,
\begin{equation}
H(\eta^{p}|\eta^{o})= \frac{p(\eta^{p}) q_{\eta^{p}}(\thick\gamma
) q(\eta^{o}|\eta^{p})}{p(\eta^{o}) q_{\eta^{o}}(\thick\gamma)
q(\eta^{p}|\eta^{o})} \bigg /
\frac{Z_{\eta^{p}}}{Z_{\eta^{o}}}   ,
\end{equation}
with $\eta^{o}$ the current value for $\eta$, one needs to take into
account that $Z_{\eta^{p}}/Z_{\eta^{o}} \neq1$. Following \citet
{moller2006}, we introduce an
auxiliary variable $w$, defined on the same state space as that of
$\thick\gamma$, which has conditional density
$f(w|\eta,\thick\gamma)$, and consider the posterior
$p(\eta, w|\thick\gamma) \propto f(w|\eta,\thick\gamma) p(\eta)
q_{\eta} (\thick\gamma) / Z_{\eta}$,
which of course still involves the unknown $Z_\eta$. Obviously,
marginalization over $w$ of $p(\eta, w|\thick\gamma)$ gives the
desired distribution $p(\eta|\thick\gamma)$. Now, if $(\eta^{o},
w^{o})$ is the current state of the algorithm, we first propose
$\eta^{p}$ with density $q(\eta^{p}|\eta^{o})$, then $w^{p}$ with density
$q(w^{p}|w^{o},\eta^{p},\eta^{o})$. As usual, the choice of these
proposal densities is arbitrary from the point of
view of the equilibrium distribution of the chain of $\eta$ values.
The choice of $f(w|\eta,\thick\gamma)$ is also arbitrary. The
key idea of the method proposed by \citet{moller2006} is to take the
proposal density for the auxiliary variable $w$ to be
of the same form as~(\ref{moller1}), but dependent on $\eta^{p}$
rather than $\eta^{o}$, that is,
\begin{equation}
\label{moller2}
q(w^{p}|w^{o},\eta^{p},\eta^{o}) = p(w^{p}|\eta^{p}) = q_{\eta^{p}}
(w^{p}) / Z_{\eta^{p}} .
\end{equation}
Then the Metropolis--Hastings ratio becomes
%
\begin{equation}
H(\eta^{p},w^{p}|\eta^{o},w^{o}) = \frac{f(w^{p}|\eta^{p},\thick
\gamma) p(\eta^{p}) q_{\eta^{p}}(\thick\gamma)
q_{\eta^{o}}(w^{o}) q(\eta^{o}|\eta^{p})}{f(w^{o}|\eta^{o},\thick
\gamma) p(\eta^{o}) q_{\eta^{o}}(\thick\gamma)
q_{\eta^{p}}(w^{p}) q(\eta^{p}|\eta^{o})},
\end{equation}
and no longer depends on $Z_{\eta^{p}}/Z_{\eta^{o}}$. The new value
$w^{p}$ for the auxiliary variable $w$ is drawn
from (\ref{moller2}) by perfect simulation using the algorithm
proposed by \citet{propp1996}.
\item[(III)] In the case of classification or survival outcomes, the
augmented data $Z$ need to be updated from their full
conditionals using Gibbs sampling; see \citet{sha2004}, \citet
{sha2006} and \citet{deukwoo2007} for details.
\end{enumerate}

\subsection{Posterior inference}\label{sec3.3}
The MCMC procedure results in a list of visited models with included
pathways indexed by $\thick\theta$ and selected genes
indexed by~$\thick\gamma$, and their corresponding relative posterior
probabilities. Pathway selection can be based on the
marginal posterior probabilities $p(\theta_k|\mathbf  T, Y)$. A~simple
strategy is to compute Monte-Carlo estimates by counting
the number of appearances of each pathway across the visited models.
Relevant pathways are identified as those with largest marginal
posterior probabilities. Then relevant genes from these pathways are identified
based on their marginal posterior probabilities conditional on the
inclusion of a pathway of interest,
$p(\gamma_j|\mathbf  T, Y, I{\{\theta_ks_{kj} =1\}})$. An alternative
inference for gene selection is to focus on a subset of
pathways, $\mathcal{P}$, and consider the marginal posterior
probability conditional on at least one pathway the gene belongs
to being represented in the model, $p(\gamma_{j}|\mathbf  T, Y, I{\{\sum
_{k \in\mathcal{P}} \theta_k s_{kj} > 0\}})$.
We note that Rao--Blackwellized estimates have been employed in
standard linear regression models, in place of frequency estimates, by
averaging the full conditional posterior probabilities of the inclusion
indicators.
These estimates are computationally quite expensive, though they may
have better precision, as noted by \citet{guan2009}.
Because of our strategy for inference, that selects first pathways and
then genes conditional on selected pathways,
Rao--Blackwellized estimates of marginal probabilities may not be
straightforward to derive. In all simulations and examples reported in
this paper we have obtained satisfactory results by simply estimating
the marginal
posterior probabilities with the corresponding relative frequencies of
inclusion in the visited models.

Inference for a new set of observations, $(\mathbf  X_f, Y_f)$, can be
done via least squares prediction,
$\widehat{Y}_f = \mathbf 1_n \tilde{\alpha} + \mathbf  T_{f(\theta,
\gamma)} \tilde{\thick\beta}_{(\theta, \gamma)}$,
where $\mathbf  T_{f(\theta, \gamma)}$ is the first principal component
based on selected genes from relevant pathways and
where $\tilde{\alpha} = \bar Y$ and $\tilde{\thick\beta}_{(\theta
, \gamma)} = (\mathbf  T_{(\theta, \gamma)}' {\mathbf
T}_{(\theta, \gamma)} + h^{-1} \mathbf  I_{K_{\theta}})^{-1} {\mathbf
T}_{(\theta, \gamma)}' Y$,
with $Y$ the response variable in the training and $\mathbf  T_{(\theta,
\gamma)}$ the scores obtained from the training data
using selected pathways and genes included in the model. Note that for
prediction purposes, since we do not know the future
$Y_f$, a PLS regression cannot be fit. Therefore, we generate
$T_{f(\theta,\gamma)}$ by considering the first latent
component obtained by applying PCA to each selected pathway using the
included genes.

In the case of categorical or censored survival outcomes, the sampled
latent variables $Z$ would be used to estimate
$\widehat Z_f$, then the correspondence between $Z$ and the observed
outcome outlined in Section \ref{sec:surv} can be
invoked to predict $Y_f$ [\citeauthor{sha2004} (\citeyear{sha2004,sha2006}); \citet{deukwoo2007}].

\section{Application}\label{sec4}
We assess performances on simulated data, then illustrate an
application to microarrays using the
KEGG pathway database to define the MRF.

\subsection{Simulation studies} \label{sec:simul}\label{sec4.1}
We investigated the performance of our model using simulated data based
on the gene-pathway relations, $\mathbf{S}$, and gene
network, $\mathbf{R}$, of 70 pathways and 1,098 genes from the KEGG
database. The relevant pathways were defined by selecting 4
pathways at random. For each of the 4 selected pathways, one gene was
picked at random and its direct neighbors that belong
to the selected pathways were chosen. This resulted in the selection of
4 pathways and 15 genes: 7 out of 30 from the first
pathway, 3 out of 35 from the second, 3 out of 105 from the third, and
2 out of 47 from the fourth pathway. Gene expressions
for $n=100$ samples were simulated for these 15 genes using an approach
similar to \citet{li2008}. This was accomplished by
first creating an ordering among the 15 selected genes by changing the
undirected edges in the gene networks into directed
edges. The first node on the ordering, which we denote by $X_{F_1}$,
was selected from each pathway and drawn from a
standard normal distribution; note that this node has no parents. Then
all child nodes directly connected only to $X_{F_1}$
and denoted by $X_{F_2}$ were drawn from $X_{F_2} \sim\mathcal
{N}(X_{F_1} \rho, 1)$.
Subsequent child nodes at generation $j$, $X_{F_j}$, were drawn using
all parents from
$X_{F_j} \sim\mathcal{N}( \rho\mathbf{X}_{pa(F_{j})} \mathbf
1_{|pa(F_j)|}, 1)$,
where $pa(F_j)$ indicates the set of parents of node $j$ and $\mathbf{X}_{pa(F_{j})}$ is a matrix containing the expressions of
all the $|pa(F_j)|$ parents for node $j$. The expression levels of the
remaining 1,073 genes deemed irrelevant were simulated
from a standard normal density. The response variables for the $n=100$
samples were generated from
\[
Y_i = \sum_{j=1}^{15} X_{ij} \beta+ \varepsilon_i,  \qquad  \varepsilon
_i \sim\mathcal{N}(0,1),   i=1, \ldots,100.
\]
For the first data set we set $\beta= \pm0.5$, with the same sign for
genes belonging to the same pathways. For the second and third data
sets we used $\beta= \pm1$ and $\beta= \pm1.5$, respectively. Note
how the generating process is
different from model (\ref{eq:model2}) being fit.

We report results obtained by choosing, when possible, hyperparameters
that lead to weakly informative prior
distributions. A vague prior is assigned to the intercept $\alpha$ by
setting $h_0$ to a large value tending to
$\infty$. For $\sigma^2$, the shape parameter can be set to $\nu
_0/2=3$, the smallest integer such that the variance of the
inverse-gamma distribution exists, and the scale parameter~$\nu
_0\sigma_0^2/2$ can be chosen to yield a weakly
informative prior. For the vector of regression coefficients, $\beta
_k$, we set the prior mean to $\beta_0 = 0$ and choose~$h$ in the range of variability of the covariates, as suggested in
Section \ref{sec:priorbeta}. Specifically, we set $h_0 =
10^6$, $\alpha_0 = \beta_0 = 0$, $\nu_0 \sigma_0/2 = 0.5$, and
$h=0.02$. For the pathway selection indicators, $\theta_k$,
we set $\varphi_k^* = 0.01$. As for the prior at the gene level, we
set $\mu=-3.5$, corresponding to setting the
proportion of genes expected a priori in the model to, at least,
3\% of the total number of genes. Parameters
$\varphi_k^*$ and $\mu$
influence the sparsity of the model and consequently the magnitude of
the marginal posterior probabilities. Some
sensitivity is, of course, to be expected. However, in our simulations
we have noticed that the ordering of
pathways and genes based on posterior probability remains roughly the
same and, therefore, the final selections are
unchanged as long as one adjusts the threshold on the posterior probabilities.
Also, for the hyperprior on $\eta$, we set $\eta_{PT}= 0.092$, to
avoid the phase transition problem,
and $c_0 = 5$ and $d_0=2$, to obtain
a prior distribution that favors bigger values of $\eta$ in the
interval $0 \leq\eta\leq\eta_{PT}$. In our
simulations we did not notice sensitivity to the specification of $c_0$
and $d_0$.

\begin{figure}

\includegraphics{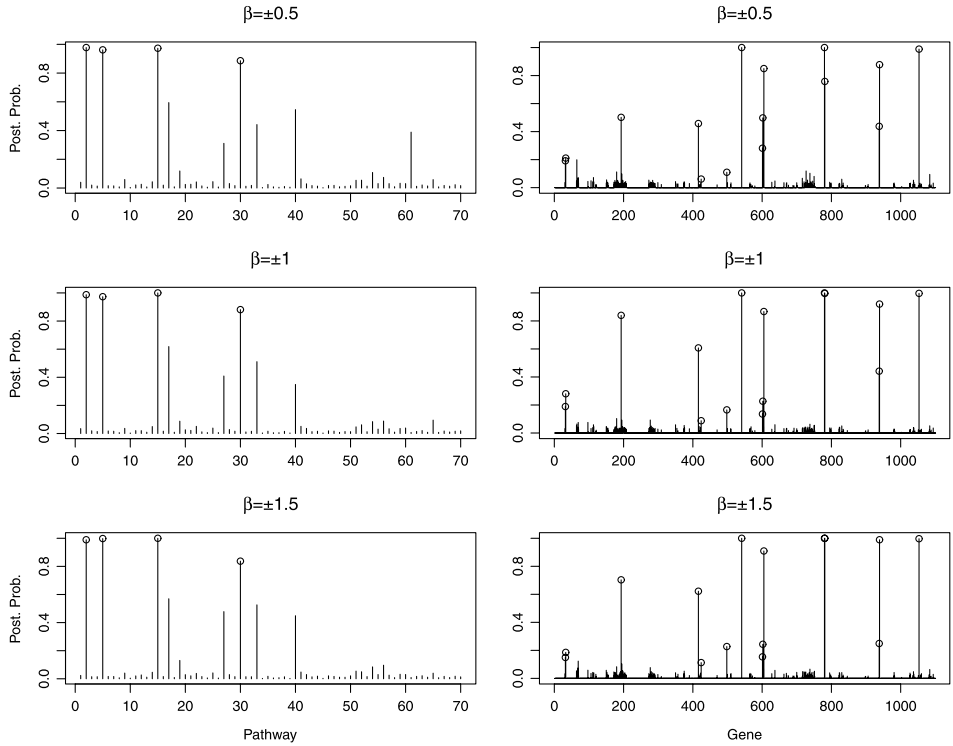}

\caption{Simulated data: Marginal posterior probabilities for pathway
selection, $p(\theta_{k}|\mathbf  T, Y)$, and conditional posterior
probabilities for gene selection, $p(\gamma_j|\allowbreak\mathbf  T, Y,\allowbreak I{\{\sum
_{k \in\mathcal{P}} \theta_k s_{kj} > 0\}})$, for the three
simulated data sets. Open circles indicate pathways and genes used to
generate the outcome variable.}
\label{fig:PPsimPathways}
\end{figure}

The MCMC sampler was run for 300,000 iterations with the first 50,000
used as burn-in. We computed the marginal posterior
probabilities for \mbox{pathway} selection, $p(\theta_k=1|Y, \mathbf  T)$, and
the conditional posterior probabilities for gene
selection given a subset of selected pathways, $p(\gamma_{j}|\mathbf  T,
Y, I{\{\sum_{k \in\mathcal{P}} \theta_k s_{kj} >
0\}})$. Figure \ref{fig:PPsimPathways} displays the marginal posterior
probabilities of inclusion for all 70 pathways and
the conditional posterior probabilities of inclusion for all 1,098 genes.

Important pathways and genes can be selected as those
with highest posterior probabilities. For example, in all 3 scenarios
all four relevant pathways were selected with a marginal posterior
probability cutoff of 0.8.
Reducing the selection threshold to a marginal posterior probability of
0.5 pulls in two
false positive pathways, for all the three simulated scenarios considered.
One of the false positives is the pathway with index 17 in Figure
\ref{fig:PPsimPathways}, which contains more than 100 genes. A~closer
investigation of the MCMC output reveals that
different subsets of its member genes are selected whenever it is
included in the model, resulting in a high marginal
posterior of inclusion for the pathway but low marginal posterior
probabilities for all its member genes. The second false
positive pathway appears to be selected often because it contains two
or three of the relevant genes that were used to
simulate the response variable and were also included in the model with
high marginal posterior probabilities; all its other
member genes have very low probabilities of selection. As expected, the
identification of the relevant genes is easier when
the signal-to-noise ratio is higher. Conditional upon the best 4
selected pathways, a marginal posterior probability cutoff of 0.5 on
the marginal probability of gene inclusion leads to the selection of 7,
8 and~8 relevant genes, for the three scenarios, respectively, and no
false positives.
With a marginal probability threshold of 0.1, 14 of the relevant genes
are selected with 4 false positives for the scenario with $\beta= \pm
0.5$, while
13 relevant genes are selected with only two false positives for the
simulated data with $\beta= \pm1$. In the simulated setting with
$\beta= \pm1.5$ all the 15 relevant genes are selected without any
false positive at a threshold of 0.12.

Generally speaking, the effect of the MRF prior depends on the
concordance of the prior network with the data. For the
simulated data, we found that the model with the MRF prior, compared to
the same model without the MRF, performs better in
terms of pathway selection, as it provides a~clearer separation between
relevant and nonrelevant pathways. In particular,
the average difference, over the three scenarios, between the relevant
pathway with the lowest posterior probability and the
nonrelevant pathway with the highest posterior probability is 0.28,
while without the MRF prior it is only 0.18.
In addition, we have observed increased sensitivity of the MRF prior in
selecting the true variables. For example, for the simulated case with
$\beta\pm1.5$, in order to select all 15 relevant genes, the marginal
probability cutoff must be reduced to 0.088 at the expense of including
3 false positives.
Other authors have reported similar results [\citet{zhang2010}]. In
the real data application we describe below, employing
information on gene--gene networks aids the interpretation of the
results, in particular, for those selected genes that are
connected in the MRF, and improves the prediction
accuracy.\looseness=-1

\subsection{Application to microarray data} \label{sec:app}\label{sec4.2}

We consider the \citet{veer2002} breast cancer microarray
data.\footnote{Available at
\href{http://www.rii.com/publications/2002/vantveer.htm}
{www.rii.com/publications/2002/vantveer.htm}.}
 Gene expression measures
were collected on each patient using DNA microarray with 24,481 probes.
Missing expressions were imputed using a $k$-nearest neighbor algorithm
with $k=10$. The procedure consists of identifying the $k$ closest
genes to the one with missing expression in array $j$ using the other
$n-1$ arrays, then imputing the missing value by the average expression
of the $k$ neighbors [\citet{troy2001}]. We focus on the 76 sporadic
lymph-node-negative patients, 33 of whom developed distant metastasis
within 5 years; the remaining 43 are viewed as censored cases. We
randomly split the patients into a training set of 38 samples and a
test set of the same size using a fairly balanced split of
metastatic/nonmetastatic cases.
The goal is to identify a subset of pathways and genes that can predict
time to distant metastasis.

The gene network and pathway information were obtained from the KEGG
database. This was accomplished by mapping probes to pathways using the
links between pathway node identifiers and LocusLink ID. 
Using the R package \textit{KEGGgraph} [\citet{jitao}], we first
downloaded the gene network for each pathway, then merged all networks
into a single one with all genes. A~total of 196 pathways and 3,592
probes were included in the analysis, with each pathway containing
multiple genes and with most genes associated with several pathways.

We ran two MCMC chains with different starting numbers of included
variables, 50 and 80, respectively. We used 600,000 iterations with a
burn-in of 100,000 iterations. We incorporated the first latent vector
of the PLS for each pathway into the analysis as described in Section
\ref{sec:reg} and set the number of pathways expected   a priori
in the model to $10\%$ of the total number. For the gene selection, we
set the hyperparameter of the Markov random field to $\mu= -3.5$,
indicating that   a priori at least 3\% of genes are expected to be
selected. We set $\eta_{PT}= 0.09$, to avoid the phase transition
problem, and $c_0 = 1$ and $d_0=1$, to obtain a noninformative prior
distribution. A sensitivity analysis showed that the posterior
inference is not affected by different values of $c_0$ and $d_0$. We
set $\alpha_0=\beta_0=0, h_0=10^6$ and $h=0.1$ for the prior on the
regression parameters and obtained a vague prior for $\sigma^2$ by
choosing $\nu_0/2=3$ and $\nu_0 \sigma^2_0/2=0.5$.

The trace plots for the number of included pathways and the number of
selected genes showed good mixing
(figures not shown). The MCMC samplers mostly visited models with 20--45
pathways and 50--90 genes. To assess the agreement of the results
between the two chains, we looked at the correlation between the
marginal posterior probabilities for pathway selection, $p(\theta_k|
\mathbf  T, Y)$, and found good concordance between the two MCMC chains
with a correlation coefficient of 0.9933. Concordance among the
marginal posterior probabilities was confirmed by looking at a scatter
plot of $p(\theta_k|\mathbf  T, Y)$ across the two MCMC chains (figure
not shown).

\begin{figure}

\includegraphics{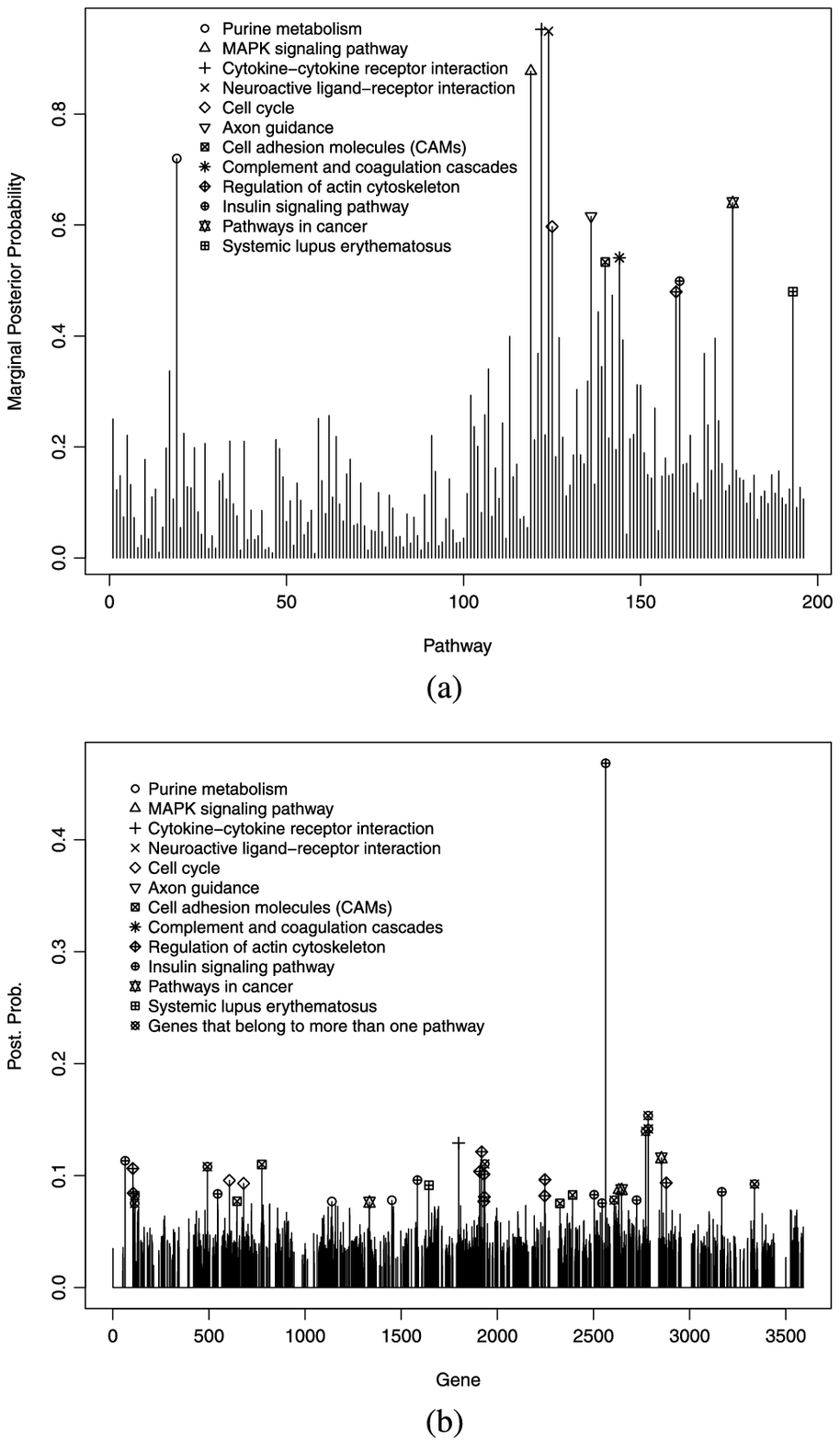}

\caption{Microarray data: Plot (\textup{a}): Marginal posterior probabilities
for pathway selection, $p(\theta_{k}|\mathbf  T, Y)$. The 12 pathways
with largest probabilities are marked with symbols. Plot (\textup{b}):
Conditional posterior probabilities for gene selection, $p(\gamma
_j|\mathbf  T, Y, I{\{\sum_{k \in\mathcal{P}} \theta_k s_{kj} > 0\}
})$. The 41 probes with largest probability that belong to the 12
selected pathways in plot (\textup{a}) are marked with $\Delta$.}
\label{fig:postgene}
\end{figure}

The model also showed good predictive performance. \citet{sha2006}
already analyzed these data using an AFT model with 3,839 probes as
predictors and obtained a predictive MSE of 1.9317 using the 11 probe
sets with highest marginal probabilities. Our model incorporating
pathway information achieved a predictive MSE of 1.4497 on the
validation set, using 12 selected pathways and 41 probe sets with
highest posterior probabilities. The selected pathways and genes are
clearly indicated in the marginal posterior probability plots displayed
in Figure \ref{fig:postgene}. If we increase the marginal probability
thresholds for selection and consider a model with 7 selected pathways
and 14 genes, to make the comparison more fair with the results of
\citet{sha2006}, we obtain a MSE of 1.7614.
As a reminder, our model selects relevant pathways and relevant genes
simultaneously, while the model of \citet{sha2006} selects genes only.
Of course, one can always select pathways post-hoc, as those that
contain the selected genes. However, as single genes belong to multiple
pathways, we expect our approach to give a more precise selection.

\begin{figure}

\includegraphics{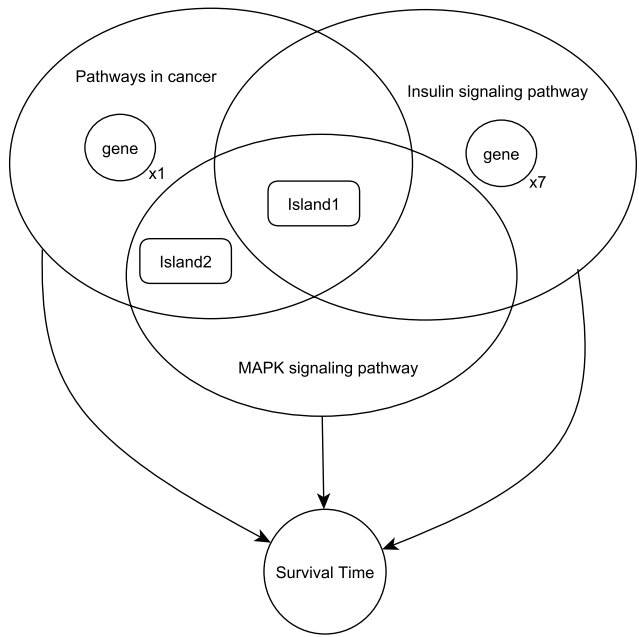}

\caption{Microarray data: Graphical representation of a subset of
selected pathways with islands and singletons. The genes in the islands
are listed in Table \protect\ref{table:islands}.}
\label{fig:finalPath}
\end{figure}

From a practical point of view, researchers can use the posterior
probabilities produced by our selection algorithm as a way to
prioritize the relevant pathways and genes for further experimental work.
For example, the genes corresponding to the best 41 selected probe
sets, conditional upon the best 12 selected pathways, are listed in
Table \ref{table:islands} divided by islands, which correspond to sets
of connected genes in the Markov random field. The islands help with
the biological interpretation by locating relevant branches of
pathways. A subset of the selected pathways along with islands and
singletons are displayed in Figure \ref{fig:finalPath}. Several of the
identified pathways are involved in tumor formation and progression.
For instance, the mitogen-activated protein kinase (MAPK) signaling
pathway, involved in various cellular functions, including cell
proliferation, differentiation and migration, has been implicated in
breast cancer metastasis [\citet{lee2007}]. The KEGG pathway in
cancers was also selected with high posterior probability. Other
interesting pathways are the insulin signaling pathway, which has been
linked to the development, progression and outcome of breast cancer,
and purine metabolism, involved in nucleotide biosynthesis and affects
cell cycle activity of tumor cells.

\begin{table}
\caption{The 41 selected genes divided by islands and with associated
pathway indices (in~parenthesis)}\label{table:islands}\label{TabPostG}
\begin{tabular}{@{\hspace*{12pt}}p{348pt}@{}}
\hline
\multicolumn{1}{@{}l@{}}{Singleton genes (no direct neighbor selected)} \\
 ACACB (10), C4A (8, 12), CALM1 (10), CCNB2 (5), CD4 (7), CDC2
(5), CLDN11 (7), FZD9 (11), GYS2 (10),
HIST1H2BN (12), IFNA7 (3), NFASC (7), NRCAM (7), PCK1 (10), PFKP (10),
PPARGC1A (10), PXN (9)\\[4pt]
\multicolumn{1}{@{}l@{}}{Island 1} \\
ACTB (9), ACTG1 (9), ITGA1 (9), ITGA7 (9), ITGB3 (9), ITGB4 (9), ITGB6
(9), ITGB8 (7, 10), MYL5 (9),
MYL9 (9), PDPK1 (10), PIK3CD (9, 10, 11), PLA2G4A (2), PLCG1 (11), PRKCA
(2, 11), PRKY (2, 10), PRKY (2, 10),
PTGS2 (11), SOCS3 (10) \\[4pt]
\multicolumn{1}{@{}l@{}}{Island 2} \\
ACVR1B (2, 3, 11), ACVR1B (2, 3, 11), TGFB3 (2, 3, 5, 11)\\[4pt]
\multicolumn{1}{@{}l@{}}{Island 3} \\
ENTPD3 (1), GMPS (1) \\
\hline
\end{tabular}
\legend{Notes: The pathway indices correspond to the
following: 1-Purine metabolism, 2-MAPK signaling pathway,
3-Cytokine--cytokine receptor interaction, 4-Neuroactive ligand-receptor
interaction, 5-Cell cycle, 6-Axon guidance, 7-Cell adhesion molecules
(CAMs), 8-Complement and coagulation cascades, 9-Regulation of actin
cytoskeleton, 10-Insulin signaling pathway, 11-Pathways in cancer,
12-Systemic lupus erythematosus.}
\vspace*{-2pt}
\end{table}

In addition, several genes with known association to breast cancer were
also selected. Protein kinase C alpha (PKCA), which belongs to the MAPK
pathway and the KEGG pathways in cancer, has been reported to play
roles in many different cellular processes, including cell functions
associated with breast cancer progression. It has been shown to be
overexpressed in some antiestrogen resistant breast cancer cell lines
and to be involved in the growth of tamoxifen resistant human breast
cancer cells [\citet{frankel}].
Patients with PKCA-positive tumors have been shown to have worse
survival than patients with PKCA-negative tumors, independently of
other factors [\citet{lonne}]. Prostaglandin-endoperoxide synthase-2
(PTGS2, also known as cyclooxygenase-2 or COX2) has also been related
to breast cancer. \citet{denkert} observed COX2 overexpression in
breast cancer and strong association with indicators of poor prognosis,
such as lymph node metastasis, poor differentiation and large tumor
size. This was further confirmed by \citet{gupta}, who showed that the
expression of COX2 in human breast cancer cells facilitates the
assembly of new tumor blood vessels, the release of tumor cells into
the circulation, and the breaching of lung capillaries by circulating
tumor cells to seed pulmonary metastasis. This is an important finding,
as the majority of breast cancer deaths result from metastases rather
than direct effects of the primary tumor. Another gene previously shown
to be predictive of breast cancer lung metastasis is integrin, beta-8
(ITGB8) [\citet{land}]. We also identified integrin, beta-4 (ITGB4)
which regulates key signaling pathways related to carcinoma
progression, and is linked to aggressive tumor behavior and poor
prognosis in certain breast cancer subtypes [\citet{guo}].

\section{Discussion}\label{sec5}

We have proposed a model that incorporates biological knowledge from
pathway databases into the analysis of DNA microarrays to identify
pathways and genes related to a phenotype. Information on pathway
membership and gene networks are used to define pathway summaries,
specify prior distributions that account for the dependence structure
between genes, and define the MCMC moves to fit the model. The gene
network prior and the synthesis of the pathway information through PLS
bring in additional information that is especially useful in microarray
data, due to the low sample size and large measurement error.
Performances of the method were evaluated on simulated data and a
breast cancer gene expression study with survival outcomes was used to
illustrate its application.

Our simulation studies show the effect of the MRF prior on the
posterior inference. In general, as expected, the effect of the prior
depends on the data and, in particular, on the concordance of the prior
network with the data. In our simulations, employing the MRF prior
allows us to achieve a better separation of the relevant pathways from
those not relevant (in particular, we have found a larger average
difference, over three scenarios, between the relevant pathway with the
lowest posterior probability and the nonrelevant pathway with the
highest posterior probability). In addition, in the simulated setting
with fairly small regression coefficients the model with the MRF prior
was able to select all the correct genes without any false positive,
while the model without MRF includes 3 false positives. Other authors
have reported improvements on selection power and sensitivity with
respect to commonly used procedures that do not use the pathway
structure, with similar, and in some cases, lower false discovery
rates. In addition, in our formulation of the model we have used
biological information not only for prior specification but also to
structure the MCMC moves. This is helpful in arriving at promising
models avoiding visiting invalid configurations. Finally, in real data
applications, we have found that employing information on gene--gene
networks can lead to the selection of significant genes that would have
been missed otherwise, aiding the interpretation of the results, and
achieving better predictions compared to models that do not treat genes
as connected elements that work in groups or pathways.

Several MRF priors for gene selection indicators have been proposed in
the literature. It is interesting to compare the parametrization of the
MRF used in this paper and in \citet{zhang2010} to the parametrization
used in \citeauthor{wei2007} (\citeyear{wei2007,wei2008}), where the prior on $\thick\gamma$
is defined as
\begin{equation}
\label{MRFweili}
P(\thick\gamma|\cdot) \propto\exp(d    n_1
- g    n_{01}),
\end{equation}
where $n_1$ is the number of selected genes and $n_{01}$ is the number
of edges linking genes with different values of $\gamma_j$, that is,
edges linking included and nonincluded genes among all pathways,
\[
n_1=\sum_{j=1}^{p} \gamma_j,  \qquad  n_{01} = \frac{1}{2} \sum
_{i=1}^p  \Biggl[ \sum_{j=1}^p r_{ij} -  \Biggl|\sum_{j=1}^p r_{ij} (1-
\gamma_i) -\sum_{j=1}^p r_{ij} \gamma_j  \Biggr|\  \Biggr].
\]
While $d$ plays the same role as $\mu$ in (\ref{PriorMRF}), the
parametrization using $g$ has a~different effect from $\eta$ on the
probability of selection of a gene. This is evident from the
conditional probability
$P(\gamma_j|\cdot, \gamma_i, i \in N_j) = \frac{\exp(\gamma_j
F(\gamma_j))}{1+\exp(F(\gamma_j))}$,
where $F(\gamma_j)= d + g \sum_{i \in N_j} (2 \gamma_i - 1)$. Higher
values of $g$ encourage neighboring genes to take on the same $\gamma
_j$ value, and, consequently, genes with nonselected neighbors have
lower prior probability of being selected than genes with no neighbors.
We felt that parametrization (\ref{PriorMRF}) was a better choice for
our purposes. First, in a context of sparsity, where only few nodes are
supposed to take value 1, a prior that assigns larger probability of
inclusion to genes with selected neighbors than to isolated genes seems
more appropriate. Second, the exact simulation algorithm of \citet
{propp1996} cannot be used to simulate from (\ref{MRFweili}). While
any other method to draw from (\ref{MRFweili}) would be acceptable, as
said by \citet{moller2006}, Markov chain methods, to sample from a
MRF, require to check at each step that the chain has converged to the
equilibrium distribution, to avoid introducing additional undesirable
stochasticity. On the other hand, one advantage of parametrization
(\ref{MRFweili}) is that no phase transition problem is associated to
the distribution.

Pathway databases are incomplete and the gene network information is
often unavailable for many genes. Thus, there may be situations where
the dependence structure and the MRF prior specification on the gene
selection indicator, $\thick\gamma$, cannot be used for all genes.
When the only information available is the pathway membership of genes,
the prior on $\thick\gamma$ could be elicited to capture other
interesting characteristics. For example, a gene can have  a
priori  higher probability of being selected when several pathways that
contain it are included in the model. We may also want to avoid
favoring the selection of a large pathway just because of its size. In
such cases, conditional on $\thick\theta$, independent Bernoulli
priors can be specified for $\gamma_j$ relating the probability of
selection to the proportion of included pathways that contain gene $j$,
adjusting for the pathway sizes, $p_k$, that is,
$\gamma_j|\thick\theta\sim  \operatorname{Bernoulli}  (c \cdot\frac
{\sum_{k=1}^K \theta_k s_{kj}/p_k }{\sum_{k=1}^K s_{kj} /p_k} )$,
with $c$ a hyperparameter to be elicited.

In our approach we have made use of PLS components as summary measures
of the expression of genes belonging to known pathways and then applied
a fully Bayesian approach for the selection of the pathways to be
included in the model, and the genes to be included within those
pathways. Penalized techniques, including lasso [\citet{tibsh1996}],
elastic net [\citet{zou2005}] and group lasso [\citet{yuan2006}] have
been studied extensively in the literature and have been successfully
applied to gene expression data. The group lasso, in particular,
defines sets of variables, then selects either all the variables in the
group or none of them. Recently, a~modification of the method was
proposed by \citet{friedman2010} using a more general penalty that
yields sparsity at both the group and individual feature levels to
select groups and predictors within each group. Our understanding of
group lasso is that the method works best in situations where variables
belonging to the same group are highly correlated, while covariates in
different groups do not exhibit high correlation. However, genes
belonging to the same pathway often do not exhibit high correlation in
their expression levels. Also, in our case there are genes belonging to
different pathways that have high correlation, as well as genes that
belong to more than one pathway. Initial investigations suggest that,
in terms of prediction MSE, Bayesian formulations of lasso methods
perform similarly to and, in some cases, better than the frequentist
lasso [see, e.g., \citet{kyung2010}]. Particularly
relevant to our approach is the work of \citet{guan2009}, who apply
Bayesian variable selection (BVS) and stochastic search methods in a
regression model for genome-wide data. In simulations they find that,
in spite of the apparent computational challenges, BVS produces better
power and predictive performance compared with standard lasso techniques.

\begin{supplement}
\stitle{Supplement}
\slink[doi]{10.1214/11-AOAS463SUPP} 
\slink[url]{http://lib.stat.cmu.edu/aoas/463/supplement.pdf}
\sdatatype{.pdf}
\sdescription{Description of the MCMC steps for $(\thick\theta
,\thick\gamma)$ and discussion on ergodicity of the Markov chain on
the restricted space.}
\end{supplement}


\printaddresses

\end{document}